\documentclass[sigchi-a]{acmart}

\usepackage{times}

\usepackage{soul}
\usepackage{url}

\usepackage{amsmath}
\usepackage{booktabs}
\usepackage[utf8]{inputenc} 
\usepackage[T1]{fontenc}    
\usepackage{booktabs}       
\usepackage{amsfonts}       
\usepackage{nicefrac}       
\usepackage{microtype}      
\usepackage{lipsum}
\usepackage{multirow}

\AtBeginDocument{%
  \providecommand\BibTeX{{%
    \normalfont B\kern-0.5em{\scshape i\kern-0.25em b}\kern-0.8em\TeX}}}

\copyrightyear{2020} 
\acmYear{2020} 
\setcopyright{rightsretained} 
\acmConference[CSCW '20 Companion]{Companion Publication of the 2020 Conference on Computer Supported Cooperative Work and Social Computing}{October 17--21, 2020}{Virtual Event, USA}
\acmBooktitle{Companion Publication of the 2020 Conference on Computer Supported Cooperative Work and Social Computing (CSCW '20 Companion), October 17--21, 2020, Virtual Event, USA}\acmDOI{10.1145/3406865.3418326}
\acmISBN{978-1-4503-8059-1/20/10}

\begin{document}

\settopmatter{printacmref=true}
\fancyhead{}

\title{Self-Evolving Adaptive Learning for Personalized Education}


\author{Junhua Liu}
\email{junhua_liu@mymail.sutd.edu.sg}
\orcid{0000-0003-4477-7439}
\affiliation{\institution{Singapore Uni. of Technology and Design}}
\additionalaffiliation{
  \institution{Forth AI}
}

\author{Lionell Loh}
\authornote{Contributed during Undergraduate Research Opportunity Program (UROP).}
\email{lionell_loh@mymail.sutd.edu.sg}
\affiliation{%
  \institution{Singapore Uni. of Technology and Design}}

\author{Ernest Ng}
\email{ernest_ng@mymail.sutd.edu.sg}
\affiliation{%
  \institution{Singapore Uni. of Technology and Design}}
\authornotemark[2]
  
\author{Yijia Chen}
\email{yijia_Chen@mymail.sutd.edu.sg}
\affiliation{%
  \institution{Singapore Univ. of Technology and Design}}
\authornotemark[2]

\author{Kristin L. Wood}
\email{kristin.wood@ucdenver.edu}
\affiliation{University of Colorado Denver}
\additionalaffiliation{%
  \institution{Singapore Uni. of Technology and Design}}
  
\author{Kwan Hui Lim}
\email{kwanhui_lim@sutd.edu.sg}
\orcid{0000-0002-4569-0901}
\affiliation{%
  \institution{Singapore Uni. of Technology and Design}}

\renewcommand{\shortauthors}{Liu and Lim, et al.}

\begin{abstract}
Primary and secondary education is a crucial stage to build a strong foundation before diving deep into specialised subjects in colleges and universities. To excel in the current education system, students are required to have a deep understanding of knowledge according to standardized curriculums and syllabus, and exam-related problem solving skills. In current school settings, this learning normally occurs in large classes of 30-40 students per class. Such a ``one size fits all'' approach may not be effective, as different students proceed on their learning in different ways and pace. To address this problem, we propose the Self-Evolving Adaptive Learning (SEAL) system for personalized education at scale. 
\end{abstract}
\begin{marginfigure}
   \centering
    \includegraphics[width=.9\linewidth]{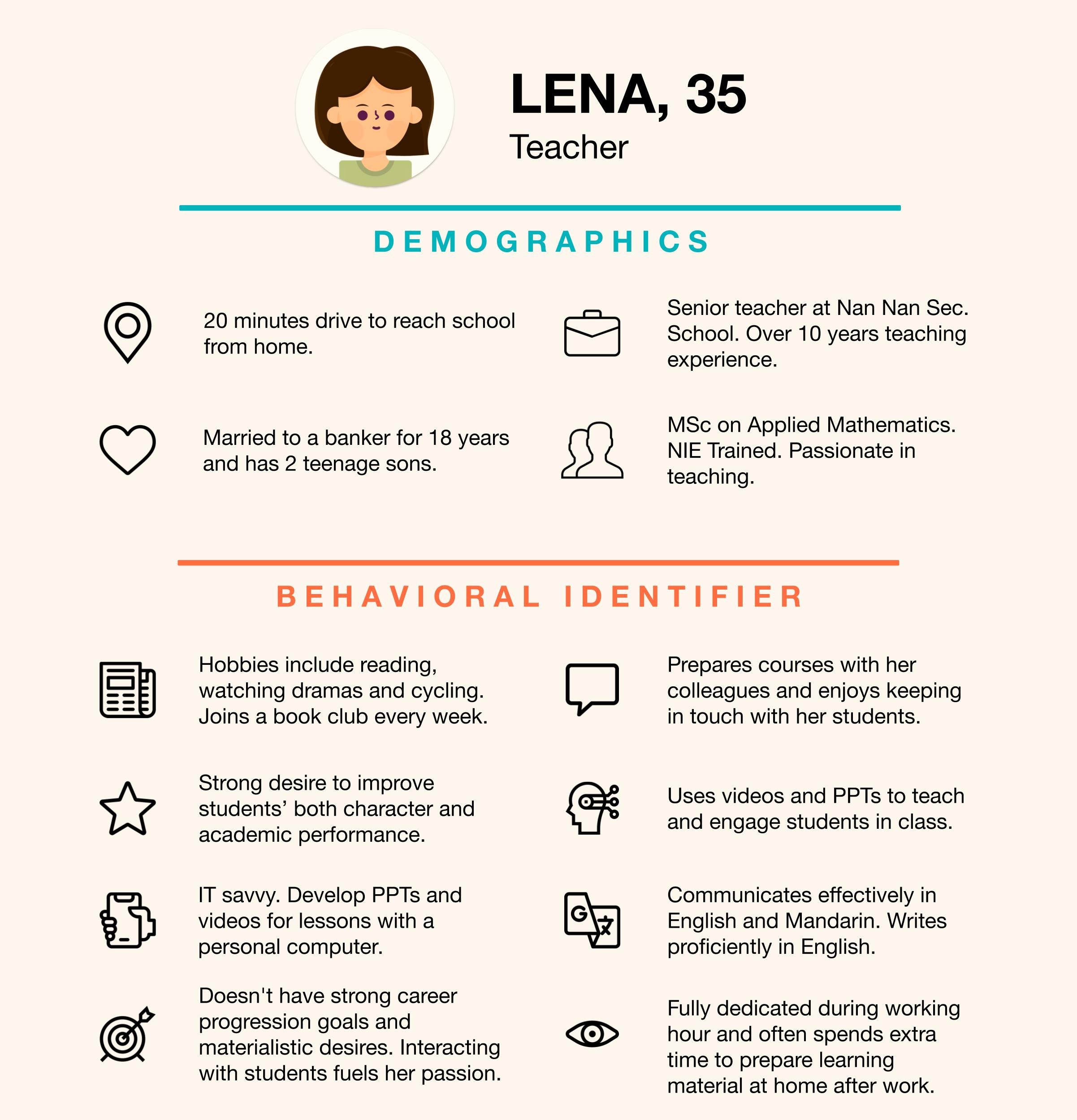}
    \includegraphics[width=.9\linewidth]{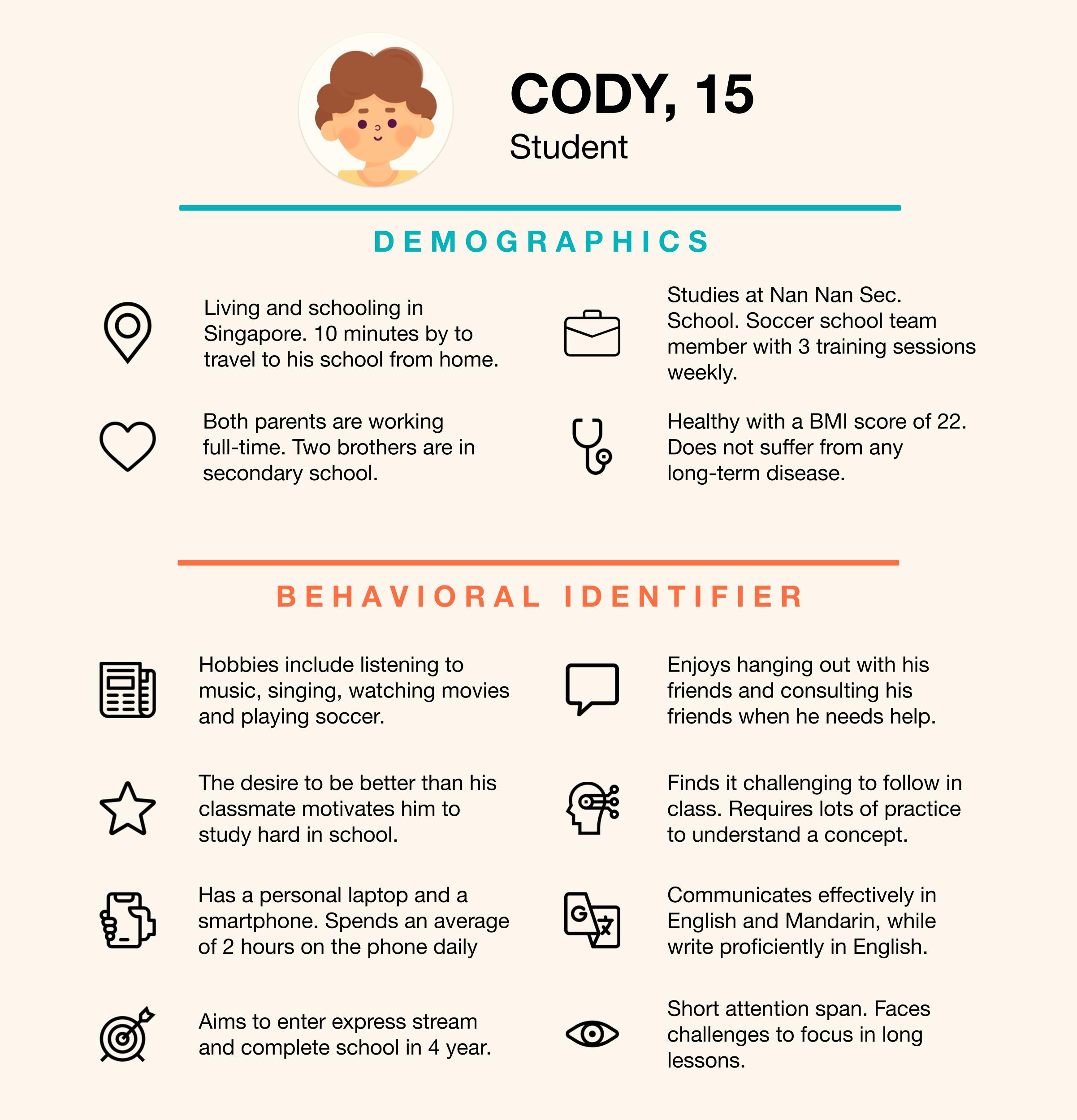}
  \caption{Personas of SEAL's main users, teachers and students}
  \label{fig:persona}
\end{marginfigure}

\maketitle

\section{Introduction}
\label{sec:introduction}

\begin{marginfigure}
   \centering
    \includegraphics[width=0.95\linewidth]{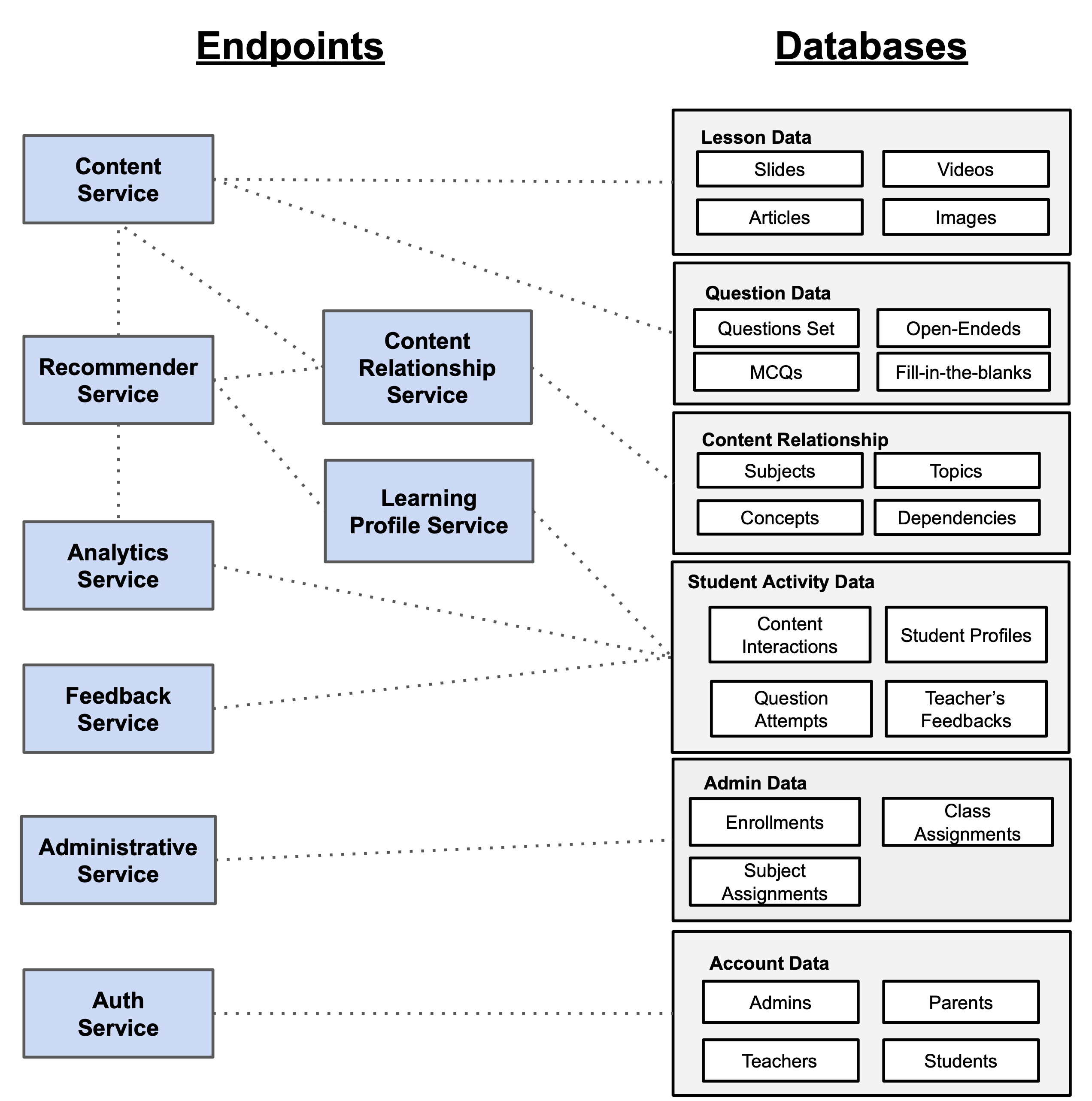}
  \caption{Server-side endpoint modules and databases}
  \label{fig:serverside}
\end{marginfigure}

Primary and secondary education serves as an important foundation for specialised subjects in colleges and universities. To excel, students are required to have a deep understanding of knowledge according to standardized curriculums and syllabus, and exam-related problem solving skills~\cite{junhua2013innovations}. However, general education has long faced the tradeoff between quality and scalability. There exists a notable advantage of small-group tuition over group instruction. Nonetheless, group instruction with large classes, i.e. over 30 students per class, remains the norm in the current school settings. As a result, the effectiveness of learning is compromised as teachers could hardly tailor personalized guidance to each student in such settings.

\begin{marginfigure}
   \centering
    \includegraphics[width=0.95\linewidth]{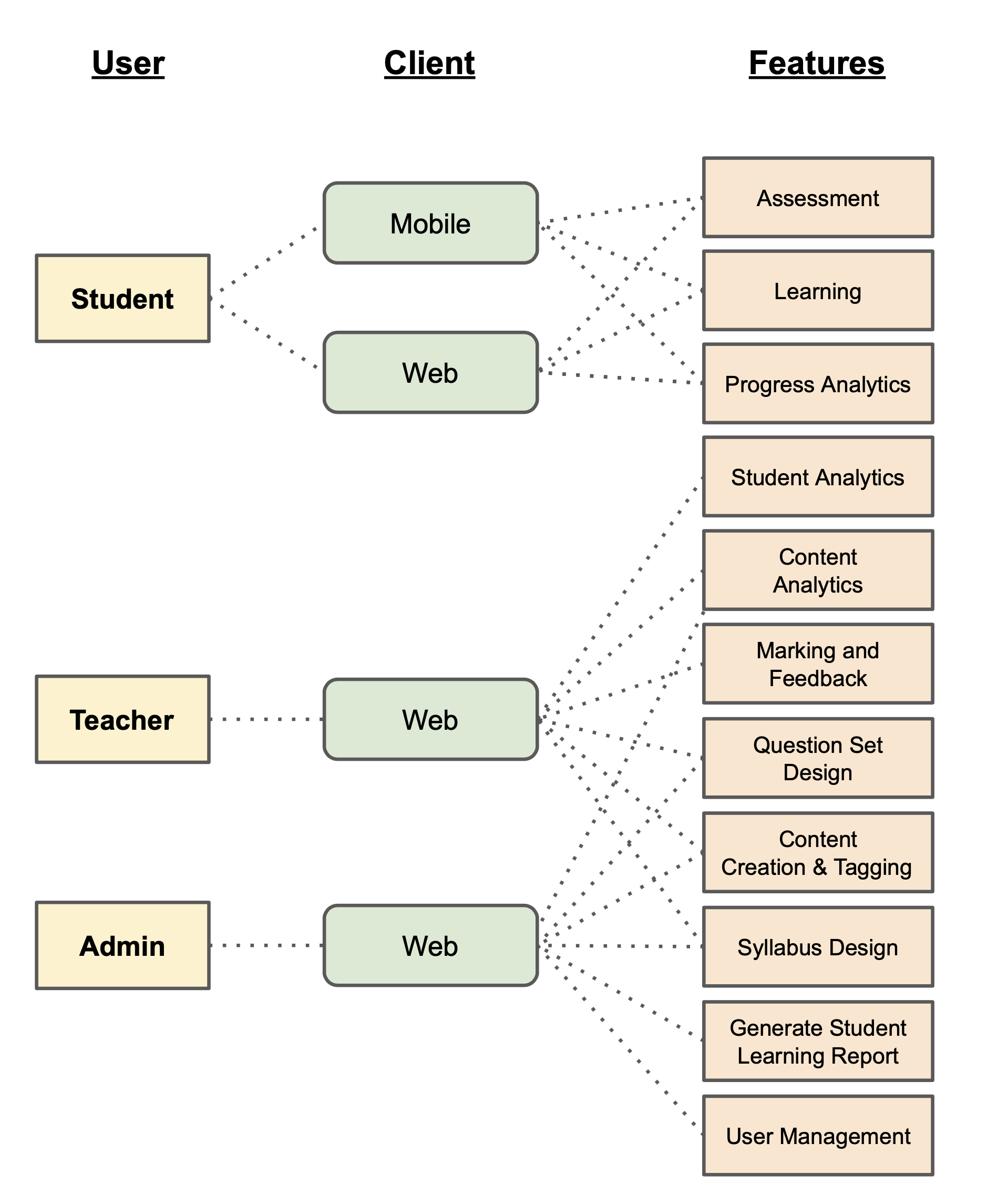}
  \caption{Client-side users, clients and key features}
  \label{fig:clientside}
\end{marginfigure}

To address this problem, we propose the Self-Evolving Adaptive Learning (SEAL) system to allow personalized learning at scale, leveraging Artificial Intelligence (AI) to solve the problem related to the trade-off between quality and scalability. SEAL aims to benefit the two most important stakeholders in the general education setting, i.e. teachers and students (Fig. ~\ref{fig:persona}), and delivers intelligent features, such as customised reports and tailor-made study guides. With AI, we solve the problem of scalability, but ensure that students still have a personalised learning experience to give them the help they need. While our use case is currently for pre-tertiary education, the SEAL system will be useful for any formal subjects or general topics.

\subsection{Significance of SEAL}

Most works on AI-enabled education focus on the automated assessment of student assignments, particularly those relating to programming exercises~\cite{blumenstein2008performance,siddiqi2010improving,naude2010marking,ala2005survey}. While assessment is an important aspect of education, existing web-based systems, such as eTutor~\cite{tekin2015etutor} and AyudasCBI~\cite{tejeda2015dynamic}, and Jill Watson SA~\cite{wang2020jill}, only measure student performance to some extent but do not personalize or influence how a student is able to acquire knowledge in specific subjects. Various works also utilize augmented reality for a more interactive learning experience~\cite{iqbal2019exploring,al2019exploring,chen2017review}.


Recently, large technology enterprises, such as Google, Microsoft and Amazon, compete in offering Machine Learning as a Service, or MLaaS~\cite{ribeiro2015mlaas,yao2017complexity}. The advent of X-as-a-service came about as internet connectivity became widely available at reasonable speeds, allowing software services to be delivered over the internet~\cite{papazoglou2003introduction}. In the same spirit, our proposed SEAL system will be offered via a similar approach. 

Our proposed SEAL system demonstrates significant advantages over existing systems in the following aspects: (i) instead of individual questions making up a subject, SEAL models the syllabus based on a knowledge graph that represents the topics, sub-topics, questions and question difficulty, which is versatile and extendable to other subjects; (ii) instead of modelling student history based on past questions answered, SEAL uses a student knowledge profile adapted from the earlier knowledge graph and the student's past performance, which allows us to better measure student competency in specific topics of that subject; (iii) SEAL offers an intuitive analytics dashboard that enables students to understand their performance and potential knowledge gaps, and track their performance over time; and (iv) SEAL adopts a service-oriented architecture that loosely couples features, endpoints and databases, which enables scalability and robustness of the system.

\begin{marginfigure}
   \centering
    \includegraphics[width=\linewidth]{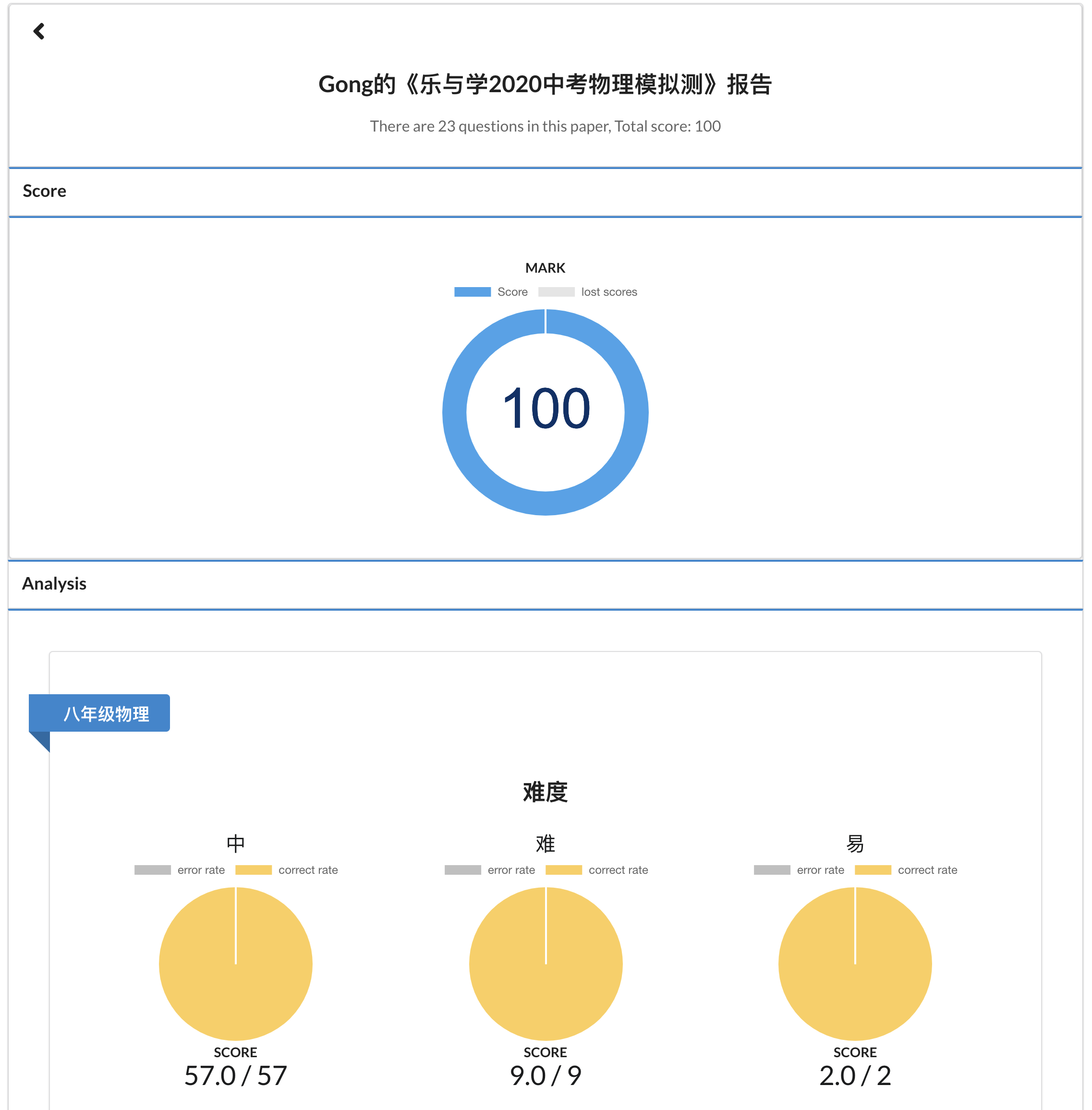}
    \includegraphics[width=\linewidth]{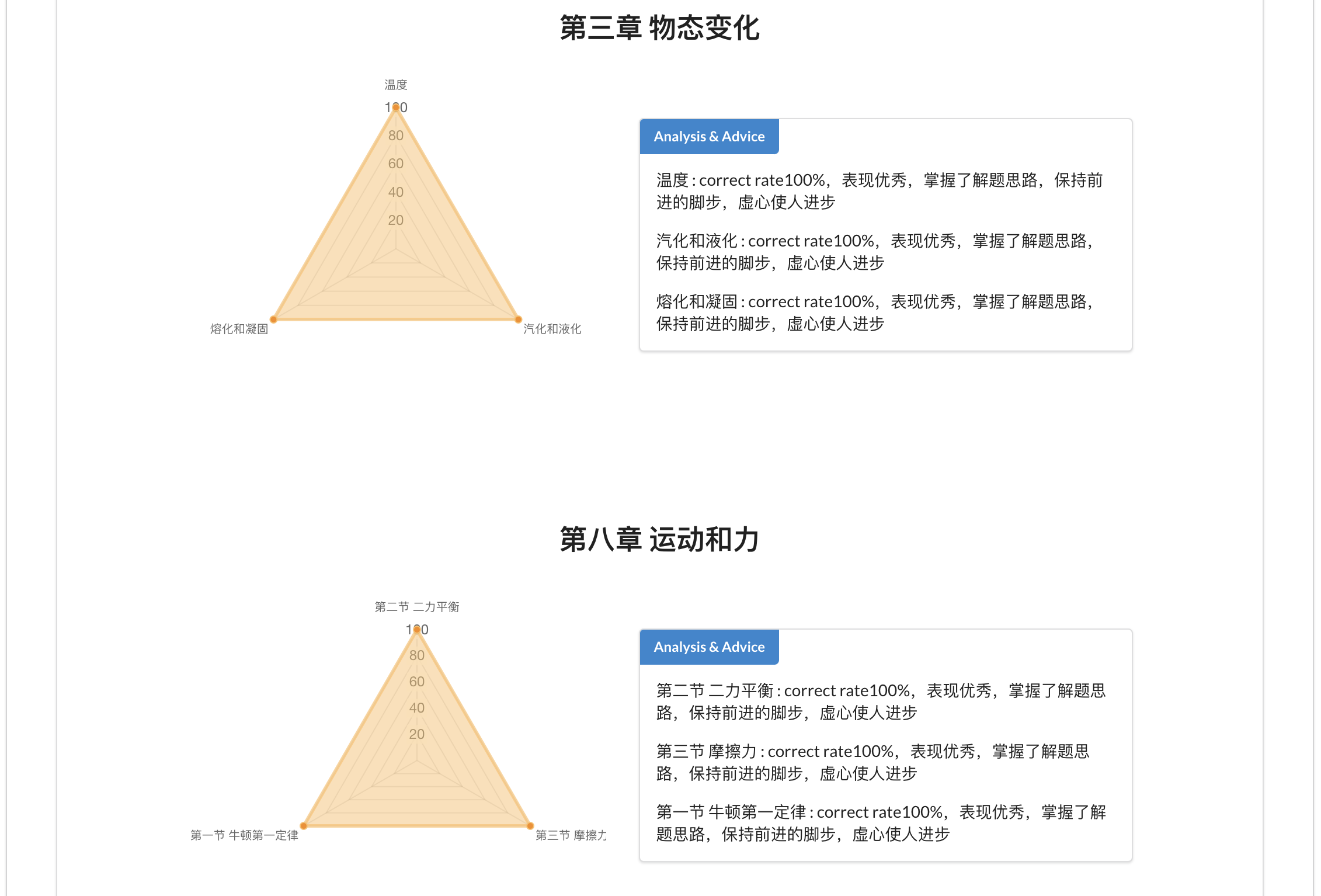}
    \includegraphics[width=\linewidth]{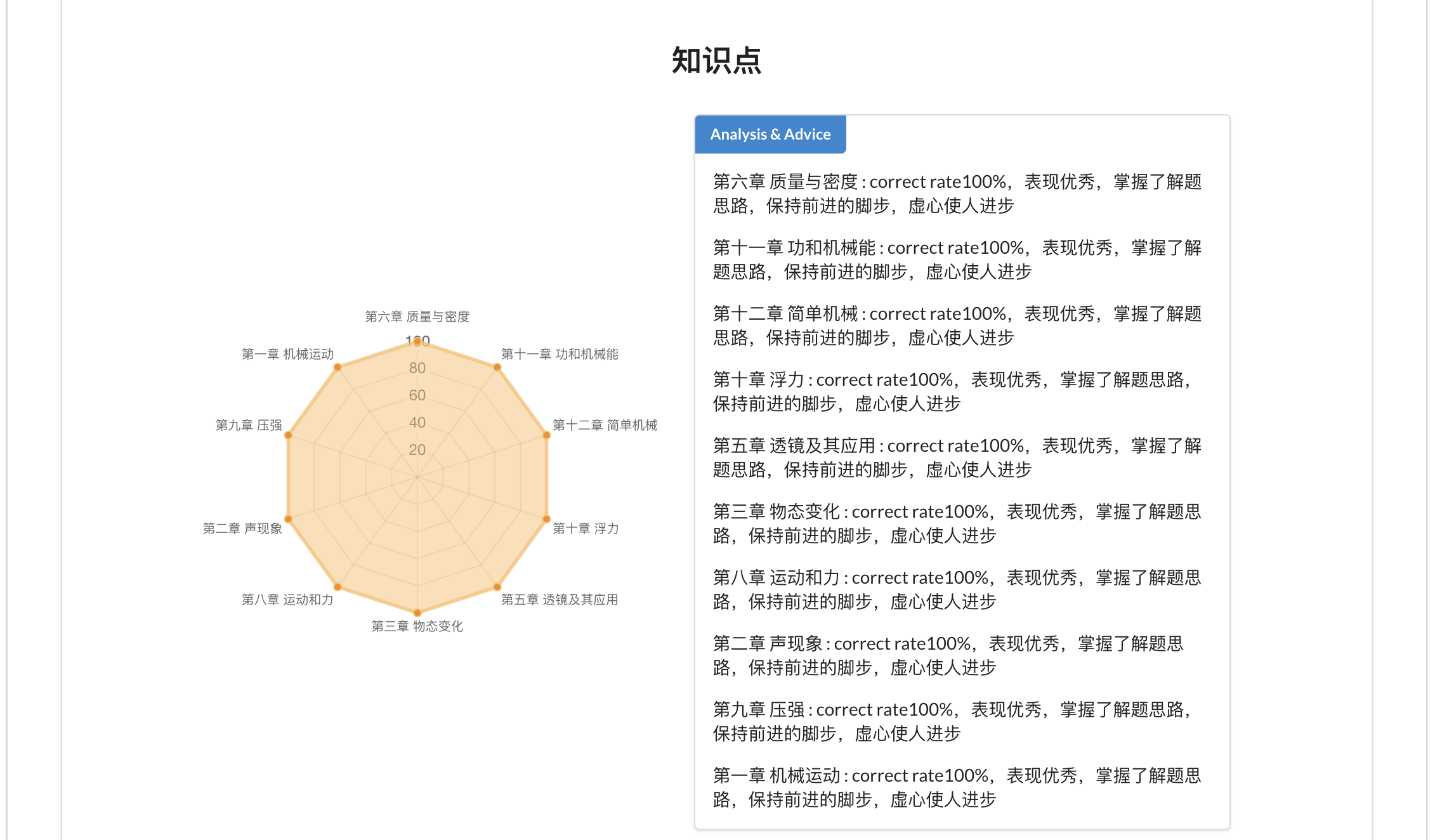}
  \caption{Analytics dashboard of SEAL}
  \label{fig:analytics}
\end{marginfigure}

\section{System Architecture}
\label{sec:system-architecture}

The system architecture can be broadly divided into two segments - server-side (Fig.~\ref{fig:serverside}) and client-side (Fig.~\ref{fig:clientside}). We adopt a Service Oriented Architecture to allow decoupling different components of the system while communicating purely via application programming interfaces (APIs). Such a loose coupling allows the system to be easily extensible to different use cases, as different web clients may choose to consume different services, and be developed independent of how the backend services or database layer have been implemented. 

\subsection{Server-side: SEALaaS}

We implement the server-side architecture as SEAL-as-a-service (SEALaaS), similarly to conventions of other MLaaS systems~\cite{ribeiro2015mlaas, papazoglou2003introduction}. Our implementation and infrastructure is designed for the wide spectrum of users to leverage our algorithms and system with varying ease and customizability. 
SEALaaS is designed to prioritise the following considerations: (i) Scalability of the solution different use cases  (ii) Usability of the system for our intended audiences. Services on their own are stateless and serve to apply computation to data Such data, which could include a student’s history of question attempts, as well as their interaction with education content, are queried from the relevant databases when a request is made. 

\subsection{Client-side: Key Features}

\begin{marginfigure}
\centering
  \includegraphics[width=0.9\linewidth]{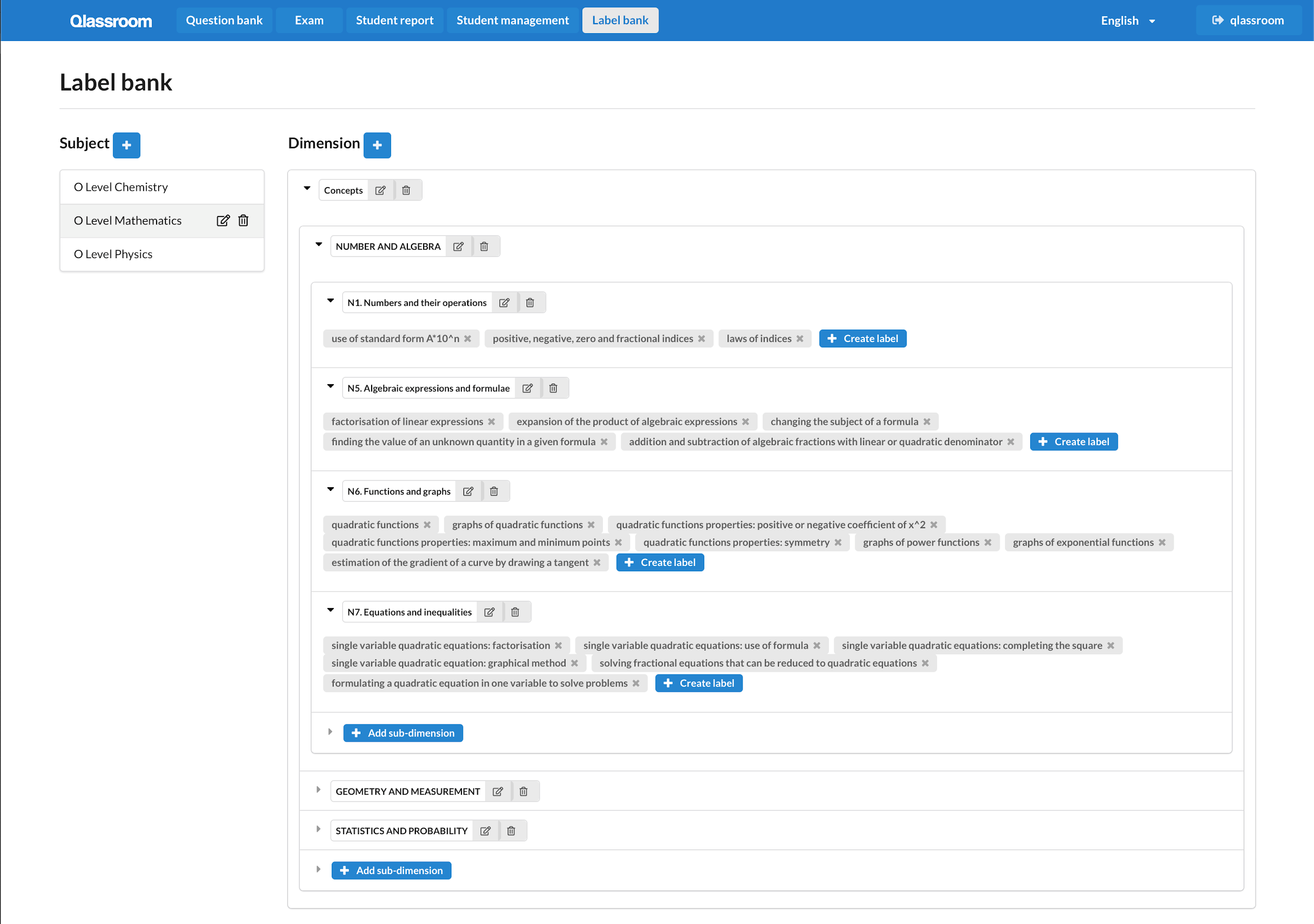}
  \caption{Teacher's tagging system.}
  \label{fig:syllabus}
\end{marginfigure}

{\bf Adaptive Assessment.}
The Adaptive Assessment feature is the core to the SEAL system. The system recommends questions to assess the student’s understanding, and learning content based on the students’ profile, learning preferences and the state of their knowledge. A comprehensive question recording and multi-dimensional tagging system (Fig.~\ref{fig:syllabus}) is implemented to assist teachers create adaptive content. Depending on what the system or user is trying to optimise for, the system will adapt to give the most efficient set of questions or answers. Some examples of optimisation objectives include the student’s mastery of difficult concepts in a subject or to optimise for a students’ confidence in their own ability. 

\begin{marginfigure}
\centering
  \includegraphics[width=0.9\linewidth]{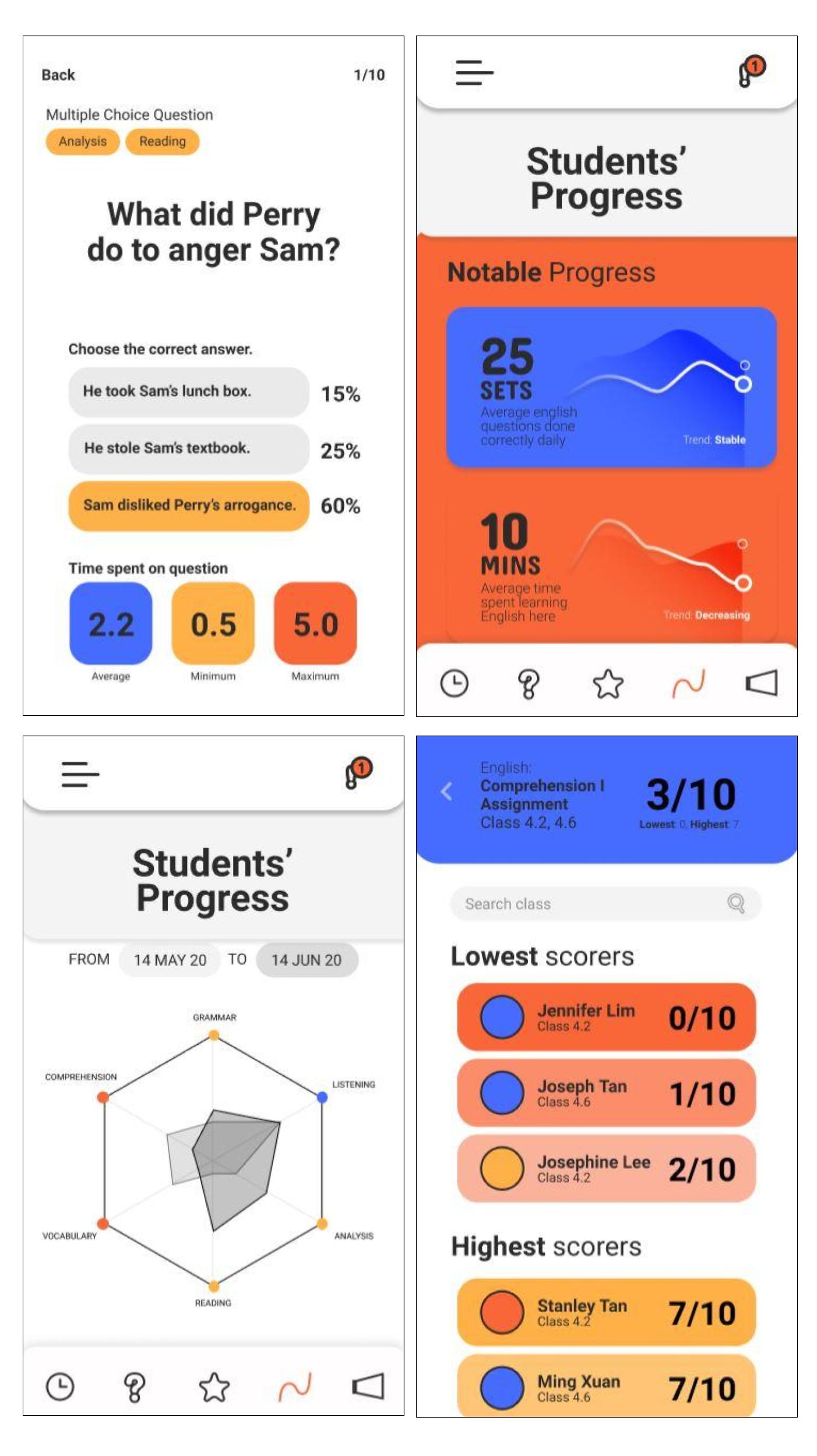}
  \caption{Mobile app user interface.}
  \label{fig:mobile}
\end{marginfigure}

{\bf Learning Analytics.}
The Learning Analytics service aggregates student activity data such as their question attempts, content interactions and teacher’s feedback to provide insight into a student’s learning progress, as well as the effectiveness of various content. The analytic reports are generated for each problem set and made accessible to both teachers and students. The forms of the reports are largely graphical,  coupled with short descriptions and suggestions that are automatically generated with templates. A sample report is shown in Fig.~\ref{fig:analytics}. The report comprises information such as the student's performance on individual quizzes, ability to answer questions of varying difficulty, identification of areas/topics the student is weak in, among other information. This component will enable students to understand their own performance and educators to track the progress of their students.

{\bf Recommender.}
The recommender service dynamically queries the Learning Profile Service, Content Service as well as the Content Relationship Service to get the most suitable set of questions or lessons for the student. The personalized and dynamic recommendation~\cite{liu2019strategic} takes into account the students’ learning preference - i.e. a student may prefer video explanation over textual, the students’ learning state - i.e. a student may have mastered first-order differentiation but not second-order differentiation as well as students’ aptitude  - i.e. a student performs well for easy questions but not for high-order questions. The recommender service takes as input what it is trying to optimise for - i.e. students’ confidence, increase in student’s aptitude or just breadth of knowledge and make a recommendation. 

\section{Conclusion and future work}

We propose the Self-Evolving Adaptive Learning (SEAL) system for personalized education, which comprises: (i) a loosely coupled SEALaaS that provides stateless endpoints; (ii) an AI-based recommender component that models students' knowledge profile, then recommends personalised content to achieve specific learning outcomes; and (iii) feature-rich client ends that facilitate teacher's teaching and student's independent learning. In future, we plan to design experiments to test the effectiveness and experience of knowledge delivery and acquisition with partners from public and private education institutions. We also intend to extend the scope of SEAL beyond general education to continuing education and training for adult learners.

\vspace{3mm}
{\noindent\bf Acknowledgements}.
This research is funded in part by the Singapore University of Technology and Design under grant SRG-ISTD-2018-140 and SUTD-UROP-1036.

\bibliographystyle{ACM-Reference-Format}
\bibliography{sample-base}

\end{document}